\documentclass[prl,aps,superbib,tightenlines,twocolumn,floatfix,letterpaper]{revtex4}
\usepackage{amsmath, graphicx, dcolumn, bm}

\begin{document}

\title{Effect of Noise on DNA Sequencing via Transverse Electronic Transport}
\author{Matt Krems$^1$, Michael Zwolak$^2$, Yuriy V. Pershin$^3$, Massimiliano Di Ventra$^1$}
\affiliation{ $^1$Department of Physics, University of California, San Diego, La Jolla, CA 92093
\\ $^2$Theoretical Division, MS-B213, Los Alamos National Laboratory, Los Alamos, NM 87545
\\ $^3$Department of Physics and Astronomy and USC Nanocenter, University of South Carolina, Columbia, SC 29208}

\date{\today}

\begin{abstract}
Previous theoretical studies have shown that measuring the
transverse current across DNA strands while they translocate through
a nanopore or channel may provide a statistically distinguishable
signature of the DNA bases, and may thus allow for rapid DNA
sequencing. However, fluctuations of the environment, such as ionic
and DNA motion, introduce important scattering processes that may
affect the viability of this approach to sequencing. To understand
this issue, we have analyzed a simple model that captures the role
of this complex environment in electronic dephasing and its ability
to remove charge carriers from current-carrying states. We find that
these effects do not strongly influence the current distributions
due to the off-resonant nature of tunneling through the nucleotides
- a result we expect to be a common feature of transport in
molecular junctions. In particular, only large scattering strengths,
as compared to the energetic gap between the molecular states and
the Fermi level, significantly alter the form of the current
distributions. Since this gap itself is quite large, the current
distributions remain protected from this type of noise, further
supporting the possibility of using transverse electronic transport
measurements for DNA sequencing.
\end{abstract}

\maketitle

\subsection{Introduction}

The prospect of sequencing an entire human genome for less than \$1000
USD in a matter of hours is becoming closer to reality~\cite{zwolak08,schloss08,branton08}.
The original DNA-nanopore experiments of Kasianowicz et al.~\cite{Kasianowicz96}
showed polynucleotides can be pulled through nanoscale pores and their
translocation detected by measuring the consequent blockage of the
ionic current through the pore. Since then, numerous experimental
studies have been performed using biological~\cite{akeson99,meller00,meller01,mathe05,butler06,astier06}
and synthetic nanopores~\cite{li03,chen04,storm05,fologea05,chang05,heng04}
which probe various physical properties of translocating polynucleotides.
This has fueled an enormous amount of research into novel sequencing
proposals based on nanopores or nanochannels~\cite{zwolak08,schloss08,branton08}.

\begin{figure}[!h]
\centering
\includegraphics[width=8cm]{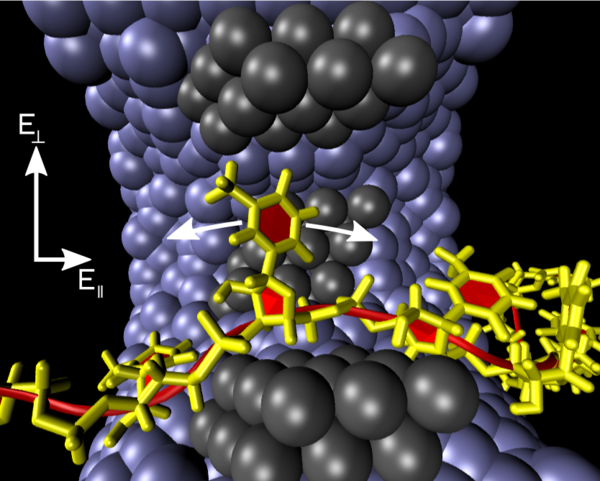}
\caption{Schematic representation of ss-DNA translocating through a
pore while the transverse electronic current is collected. The light
(purple) atoms are the silicon nitride pore and the dark (black)
atoms represent the electrode surfaces within our molecular dynamics
simulations. The single strand of DNA translocates through the pore
pulled by a longitudinal electric field, $E_{\parallel}$, and the nucleotides also experience a transverse
electric field, $E_{\perp}$. The white arrows around the DNA base
indicate an acoustic phonon-like motion which contributes to the
noise. The visualization was made with VMD~\cite{hump96}.
\label{fig:schematic}}
\end{figure}

One sequencing idea suggests detecting transverse electron currents
as single-stranded DNA (ss-DNA) translocates through a
pore~\cite{zwolak05,lagerqvist06,lagerqvist07,lagerqvist07b}.
Previous theoretical work showed the four DNA nucleotides possess
statistically distinguishable electronic
signatures~\cite{zwolak05,lagerqvist06} in the form of current
distributions when accounting for structural distortions and partial
control of the DNA dynamics (i.e., by a transverse
field)~\cite{lagerqvist06,lagerqvist07,lagerqvist07b}. These results
indicate DNA sequencing is, in principle, possible via transverse
current measurements. However, such studies have neglected
scattering processes, such as fluctuations of the environment, which
introduce electronic noise, and may thus affect the ability to
distinguish the bases.

Recently, experimentalists have successfully embedded electrodes
into solid state nanopores and
nanochannels~\cite{gierhart08,fischbein07,liang08,maleki09} and are
getting closer to measuring electronic currents with single
nucleotides present in the gap between the electrodes. When the
latter is achieved, one question which will arise is how does the
noise induced by the environment - noise which is beyond that due to
``static'' structural distortions of the nucleotides - affects the
nucleotides' electronic signatures, i.e., the current distributions.
The environment is composed of ionic and water fluctuations and
other excitations which may drastically affect the electron
dynamics, and thus the current~\cite{diventra08}. To complicate
matters, the liquid environment can scatter
electrons out of their current-carrying states by absorbing them into
the solution and allowing the longitudinal field (that pulls the DNA through the pore)
to carry them away. The influence of these and related factors can be very
important, as seen in previous studies of electronic transport
through DNA~\cite{zwolak04,endres04,shapir08}, and so far no study
has examined such effects in detail.

In this article, we address these issues theoretically. Clearly, a
fully time-dependent calculation with inclusion of all these types
of scattering processes would be ideal~\cite{diventra08}. However,
the complexity of the problem we consider, both in the
number of atoms involved and the type of scattering processes to
take into account, makes this type of dynamical calculation
unrealistic at present. Instead, we use a simplified model to
capture some of the physics we deem important and leave a
time-dependent treatment for future investigation.

In general, one expects any type of electronic noise to eventually destroy
the capability to distinguish the DNA bases once its strength is
sufficiently large. Indeed, we do find this type of
behavior. However, the noise strength at which the electronic
transport is negatively influenced is very large, beyond the
strength one would expect in realistic experimental situations. This
is due to the off-resonant nature of tunneling through the
nucleotides, and we thus expect this result to be a common feature
of molecular junctions. In other words, the separation of the energy
levels of the nucleotides from the equilibrium Fermi level
``protects'' the electronic signature of the bases. The present
study will thus help researchers understand future experimental
data, and provides further support to the viability of DNA sequencing via transverse
electronic transport.

\subsection{Setup and Methods}

As our starting point, we employ molecular dynamics simulations performed with NAMD2~\cite{phillips05} to pull
homogeneous ss-DNA through a Si$_{3}$N$_{4}$ nanopore with embedded
gold electrodes. Our basic setup is shown in Fig.~\ref{fig:schematic} and is the same as that used in previous work~\cite{lagerqvist06,lagerqvist07},
except the new trajectories here correspond to longer simulation times.
These trajectories give us the real-time atomistic structure of ss-DNA
as it propagates through the pore. With these structures, we calculate
the electronic transport in the transverse direction across the pore.
In the latter calculations, we include the effect of noise as discussed
below.

The details of the simulations are as follows. The pore is made of
2.4 nm thick silicon nitride material in the $\beta$-phase. The nanopore
hole has a double conical shape with a minimum diameter of 1.4 nm
located at the center of the membrane and an outer diameter of 2.5
nm (see Fig.~\ref{fig:schematic}). The inner diameter is chosen
wide enough such that ss-DNA is able to pass through but narrow enough
that an appreciable tunneling current can be detected. The nanopore
is then solvated in a TIP3 water sphere of 6.0 nm radius with spherical
boundary conditions in an NVT ensemble and with a 1 M solution of
potassium and chlorine ions. The CHARMM27 force field~\cite{foloppe99,mackerell99}
is used for the interaction of DNA, water, and ions, while UFF~\cite{wendel92}
parameters are used for the interaction of the Si$_{3}$N$_{4}$ membrane
and other atoms. The Si$_{3}$N$_{4}$ atoms are assumed to be fixed
during the simulation (this does not affect the conclusions). A 1
fs timestep is used and the system temperature is kept at room temperature
with a Langevin dampening parameter of 0.2 ps$^{-1}$ in the equations
of motion~\cite{aksimentiev04}. The van der Waals interactions are
gradually cut off starting at 10 \AA$\,$ from the atom until reaching
zero interaction 12 \AA$\,$ away. The energy was initially minimized
in 1000 time steps.

\begin{figure}[!h]
\centering
\includegraphics[width=8.66cm]{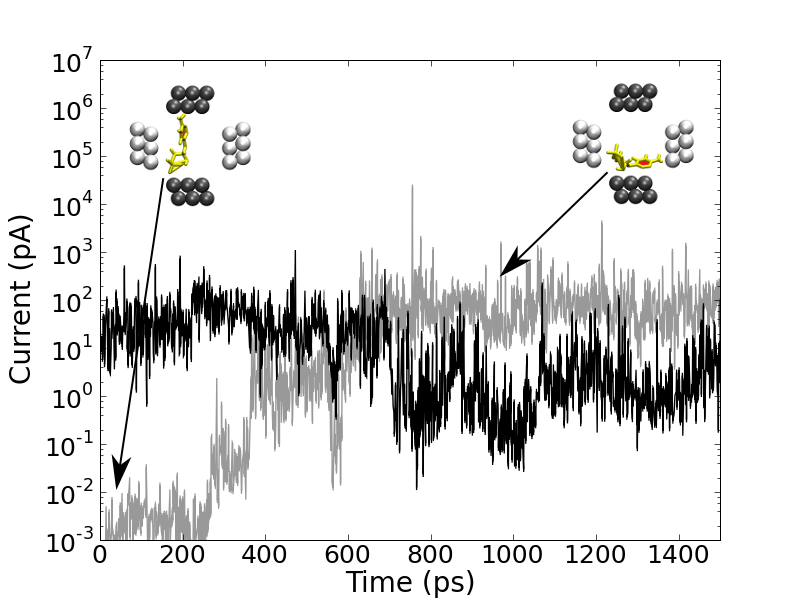}
\caption{Currents as a function of time across two pairs of
perpendicularly placed electrodes for poly(C)$_{15}$ with one base
originally aligned parallel to a pair of opposite electrodes (see
inset). The black current trace corresponds to the current from the black electrodes, and likewise the gray current trace corresponds to the gray electrodes. At time $t=0$, a bias of 1 V oriented perpendicular to the base
plane is switched on. The corresponding field aligns the
base and backbone with the gray pair of electrodes (as shown in the
inset), with a corresponding increase in the current across that pair of electrodes.
\label{alignment}}
\end{figure}

A single strand of DNA is constructed by removing one strand from a
helical, double-stranded polynucleotide created using the Nucleic
Acid Builder of the AmberTools package~\cite{macke98}. At the
initial time of the simulation, the ss-DNA is placed parallel to the
pore axis with the first base just inside the pore. The ss-DNA is
driven through the pore with a global electric field of 6 kcal/(mol
\AA e) to achieve reasonable simulation times. In the calculation of
the electronic transport, the longitudinal pulling field is turned
off and a transverse field (of the same magnitude as that driving
the current) is turned on at a moment when a base is between the
electrodes. This approximates the situation when the transverse
field is much larger than the longitudinal field. We envision this
as the typical operating regime for a sequencing device as it allows
for the suppression of a significant amount of structural
distortion~\cite{lagerqvist06}. The particular time to stop the
translocation is chosen by visual inspection. This stopping time is not particularly important because it generally takes on the
order of hundreds ps for the transverse field, $E_{\perp}$, to align
the nucleotide with the electrodes~\cite{lagerqvist07}.
Single-stranded DNA differs from double-stranded DNA in that the
persistence length of the polynucleotide is much shorter. This, in
particular, allows for the base to quickly align with the
perpendicular electric field. An example of this is reported in
Fig.~\ref{alignment} where poly(C)$_{15}$ is such that a
single C base is aligned parallel to a pair of opposite electrodes.
A bias of 1 V oriented perpendicular to the base plane is
then turned on. From the figure it is clear that, for this
particular polynucleotide and initial condition, the base and
backbone reorient themselves towards the field within about 800 ps.
This is also confirmed by the currents as a function of time across
two pairs of perpendicularly placed electrodes. At $t=0$ the largest
current is from the pair of electrodes parallel to the plane of
the base, while after 800 ps, the largest current is from the opposite
pair of electrodes. It is also evident from the figure that the
rotation does not occur uniformly in time but it proceeds by fast
rotations, followed by periods of time in which the system is
temporarily trapped in a local energy minimum. Faster rotations have
been observed with other initial conditions, transverse voltages and nucleotide
strands~\cite{lagerqvist07}, but we cannot exclude the possibility
that, for other initial conditions, longer times would be needed for
a complete rotation of the bases.

The current calculations are performed within a single-particle
scattering approach using a tight-binding Hamiltonian (see, e.g.,
Ref.~\cite{diventra08}). These calculations include water, although,
within our approach, water has little direct effect on the
current~\cite{lagerqvist07}. ``Snapshots'' of the atomistic
structure of ss-DNA between the gold electrodes are taken from the
molecular dynamics at regular time intervals. These coordinate
snapshots are used to obtain the tight-binding Hamiltonian. For each
carbon, nitrogen, oxygen, and phosphorous atoms,
$s,p_{x},p_{y},p_{z}$ orbitals are used, whereas for gold and
hydrogen only $s$ orbitals
are employed. The Fermi level is taken to be that of bulk gold
\footnote{The tight-binding Hamiltonian is constructed at every snapshot using
the YAEHMOP package (http://yaehmop.sourceforge.net/), with the Fermi
level also consistently calculated using this method.%
}.

To obtain the current across the ss-DNA, we use the retarded Green's
function, \begin{equation}
G_{DNA}(E)=\frac{1}{ES_{DNA}-H_{DNA}-\Sigma_{t}-\Sigma_{b}-\Sigma_{n}},\label{eq:retarded}\end{equation}
 where $E$ is the energy, $S_{DNA}$ and $H_{DNA}$ are the overlap
and Hamiltonian matrices, respectively, of the contents of the gap
between the electrodes (we will call it electronic junction),
$\Sigma_{t(b)}$ are the self-energy terms associated with the
interaction between the electrodes and the junction contents, and
$\Sigma_{n}$ is the self-energy associated with the noise. The
Green's function for gold needed to calculate $\Sigma_{t(b)}$ is
approximated as in Ref.~\cite{pecchia03}. We use a white-noise term,
which corresponds to a noise timescale via \begin{equation}
\tau_{n}=-\frac{\hbar}{\mathrm{Im}\{\Sigma_{n}\}},\label{eq:dep}\end{equation}
and we also take $\mathrm{Re}\{\Sigma_{n}\}=0$ (see discussion
below). This timescale sets a decay time due to interaction with the environment. The latter
can be thought of as a noise probe that
interacts with the contents of the junction~\cite{Butt, diventra08}.

If we were to follow this type of reasoning we would then set the
current in the probe to be zero and calculate the total transmission
coefficient as
\begin{equation}
T(E)=T_{tb}\left(E\right)+T_{p}\left(E\right),\label{eq:transmission}\end{equation}
where

\begin{equation}
T_{tb}\left(E\right)=\text{Tr}\left[\Gamma_{t}G_{DNA}\Gamma_{b}G_{DNA}^{\dagger}\right]
\end{equation}
is the transmission coefficient that directly couples electrodes
that measure the current in the presence of the noise probe with
$\Gamma_{t(b)}=i\left(\Sigma_{t(b)}-\Sigma_{t(b)}^{\dagger}\right)$.
This transmission contribution includes only elastic processes, as
we discuss in more detail below.

The other term is
\begin{equation}
T_{p}\left(E\right)=\frac{T_{tn}T_{nb}}{T_{tn}+T_{nb}}
\label{Tp}
\end{equation}
where
\begin{equation}
T_{\mu\nu}\left(E\right)=\text{Tr}\left[\Gamma_{\mu}G_{DNA}\Gamma_{\nu}G_{DNA}^{\dagger}\right]\label{Tn}
\end{equation}
is instead the transmission from reservoir $\mu$ to $\nu$, namely it
takes into account processes that can drive electrons out of the
electrodes into the noise probe and vice versa.

The current is then given by
\begin{equation}
I=\frac{2e}{h}\int_{-\infty}^{\infty}dET(E)\left[f_{t}(E)-f_{b}(E)\right],\label{eq:current}\end{equation}
 where $f_{t(b)}$ is the Fermi-Dirac function of the top (bottom)
electrode~\cite{diventra08}. The current distribution for a
nucleotide is the distribution obtained from the various snapshots
while the nucleotide fluctuates between the electrodes.

We will later make a microscopic connection to the above
transmission probability by starting with a Hamiltonian for
independent electrons coherently coupled to a phonon environment.
However, this analysis leads us to conclude that in the complex
liquid environment the term in Eq.~\ref{Tp} cannot correctly
represent the physical situation at hand. In fact, retention of such
term would give rise to unrealistically large currents (several
orders of magnitude larger than what we present here). While this
result would naively suggest that such currents could in fact be
observed in the present case, it is unlikely that the nanopore
environment would allow for the coherent coupling between the
electrons and excitations that gives this increased current.
Furthermore, it is likely that due to the presence of the
longitudinal field that drives the DNA through the pore the
electrons scatter out of their current-carrying states. In this work
we will then assume that current-carrying electrons can be scattered
into the complex liquid environment and retain only the first
term $T_{tb}$ in the transmission probability of Eq.~\ref{eq:transmission},
and analyze its effect as a function of the timescale strength
$\tau_{n}$. This is equivalent to assuming that the liquid
environment is represented by two probes connected to the
junction, and the probes' electrochemical potentials are adjusted so that
the combined current from the two probes into either electrode is zero.~\footnote{This condition
entails that $I_1+I_2=0$, where 1 and 2 are the two probes. This together with current conservation,
$I_t+I_b+I_1+I_2$=0, yields $I_t=-I_b$, which is the current calculated in this paper.}

\begin{figure}
\centering
\includegraphics[width=8.66cm]{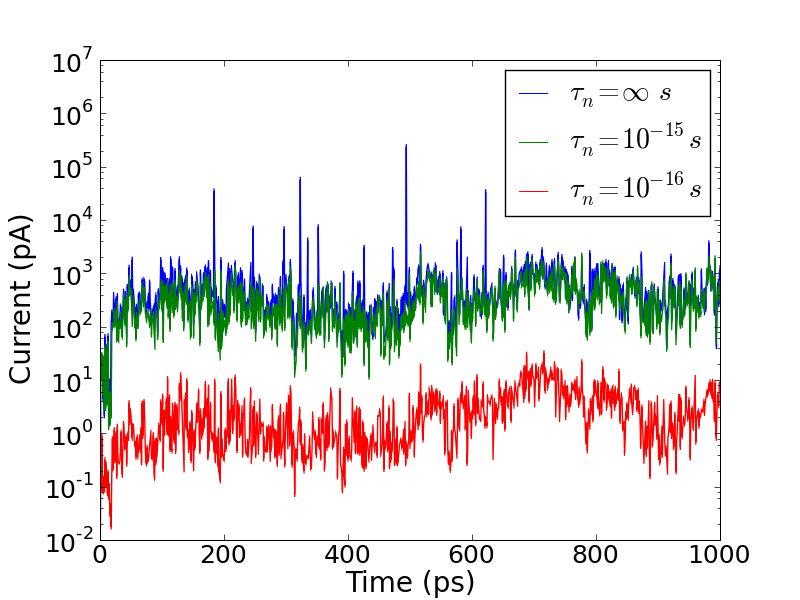}
\caption{Transverse current versus time for poly(A)$_{15}\,$ at a
transverse bias voltage of 1.0 V. The noise lowers the current
slightly for $\tau_{n}=10^{-15}\,$s. Only at the unrealistic
$\tau_{n}=10^{-16}\,$s does the current shift significantly. Slower
noise timescales give essentially the same current as the case with
an infinite noise timescale.}
\label{fig:CurrentA}
\end{figure}

{\em Noise} - As stated above, previous theoretical studies have
shown the current distributions caused by DNA static structural
distortions are statistically
distinguishable~\cite{zwolak05,lagerqvist06,lagerqvist07,lagerqvist07b}.
These studies, however, have not included the effects of external
noise. We focus specifically on noise given by Eq.~\ref{eq:dep}
because it represents many processes which happen in an experiment.
These include fast processes, such as electronic interactions with
bound waters or charges on the pore walls, and also slow processes,
such as the dynamic movement of the DNA bases and ions. From visual
inspection of the molecular dynamics simulations, we observe that
the bases fluctuate in a way reminiscent of acoustic phonons, i.e.,
we observe only low-energy excitations. An example of these
excitations is represented in Fig.~\ref{fig:schematic}, where these
slow oscillations, while not periodic, are mostly in the
longitudinal direction. No oscillations where the bases are, e.g.,
in a ``breathing mode'', that is where the base itself is expanding
and contracting, causing large energy relaxation, were observed. At
low bias, these are also unlikely to be excited by the electrical
current itself, so that we expect a low exchange of energy with the
current-carrying electrons~\cite{chenA03,chenA04}. Furthermore, we
assume the timescale for noise, Eq.~\ref{eq:dep}, is a constant for
all molecular states in the junction. In certain cases, this most
likely overestimates the effect of the noise, but, on the other
hand, it misses {}``colored noise'' effects, where, for instance,
the noise has a strong component at a particular frequency. In the
absence of a physical model for such noise which is supported by
experiments, its effect is only speculative at this stage, and we
thus defer its study for future research.

\subsection{Results and Discussion}

We have performed current calculations for some representative noise
timescales~\cite{zwolak02}: $\tau_{n}=\infty,10^{-13},10^{-14},10^{-15},10^{-16}\,$s
with transverse voltages of $0.1\,$V and $1.0\,$V. The timescale
of $10^{-16}\,$s is a particularly fast and unphysical timescale
but was used to show the onset of major differences in the current
and current distributions.

\begin{figure}
\centering
\includegraphics[width=8.66cm]{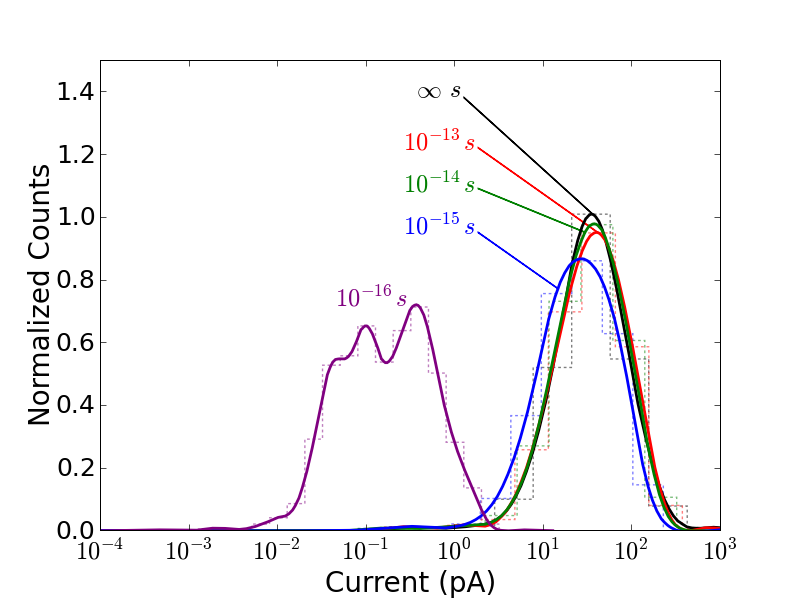}
\caption{Probability distributions for poly(A)$_{15}$ with various
noise timescales for a transverse bias voltage of 1.0 V. The very
light dashed lines correspond to the bins used to produce the
current distributions. The solid lines are interpolated from the
dashed ones. Like the current itself, only with very fast noise,
$\tau_n=10^{-16}\,$s, does the distribution change and shift
appreciably. At $\tau_n=10^{-15}\,$s, the distribution's mean shifts
slightly and it broadens somewhat.}
\label{fig:DistributionsA}
\end{figure}

For weak noise, ($\tau_{n}=10^{-13}$s - $10^{-14}$s), the average
current itself is essentially unchanged as well as the
distributions. The average percent change of an individual current
value for $\tau_{n}=10^{-13}\,$s is only about $0.1\,\%$. For
$\tau_{n}=10^{-14}\,$s, it is $1.5\,\%$. However, for a single
current count, the current may vary by orders of magnitude due to
the noise, further strengthening the argument that a single base
measurement is likely not enough to distinguish the
bases~\cite{lagerqvist06}. From Figs.~\ref{fig:CurrentA}
and~\ref{fig:DistributionsA}, $\tau_{n}=10^{-15}\,$s lowers the
current on average and slightly alters the distributions. There is
an average current reduction of about $30\,\%$. At the unphysical
fast timescale of $10^{-16}$s, the current is significantly lowered
and the distributions are pushed into an unmeasurable regime.
However, we are not aware of a physical process that may cause such
strong noise under the experimental conditions envisioned in this
work.

We have found above that even relatively strong noise does not
negatively impact the current distributions. This may seem an
unexpected result, and it will be helpful for future experimental
and theoretical efforts to understand the reason for such an effect.
We thus develop a model system to understand this behavior, as well
as the noise processes we are including. Our starting point is based
on our previous work on transverse transport through
DNA~\cite{zwolak05,lagerqvist06,lagerqvist07,lagerqvist07b}. In an
ideal configuration of a nucleotide between electrodes, the LUMO
level of the base is closest to the gold Fermi
level~\cite{zwolak05,zwolak08} and also couples well to both
electrodes. Thus, it is reasonable to treat a nucleotide in the
electronic junction as a single energy level, $E_{0}$.

At this point we may consider a model Hamiltonian
representing this level interacting with a bosonic
environment~\footnote{For simplicity, we set $\hbar=1$ in Eqs.~\ref{eq:Ham}-\ref{eq:Gapprox}.}
\begin{equation}
H=E_{0}d^{\dagger}d+H_{de}+H_{e}+d^{\dagger}d\sum_{k}g_{k}\left(b_{k}^{\dagger}+b_{k}\right)+\sum_{k}\omega_{k}b_{k}^{\dagger}b_{k},\label{eq:Ham}
\end{equation}
 where $d^{\dagger}d$ represents the occupation of the DNA LUMO level, $H_{de}$ is
the DNA-electrode interactions, and $H_{e}$ is the electrodes'
Hamiltonian. The two remaining terms represent an interaction with a
bosonic environment in the junction with interaction strength $g_k$ to
each mode $k$.
To get a tractable model, we make a few additional assumptions.
First, we assume the junction DNA energy level is equally coupled to
all levels of both electrodes and that we are at low enough bias and
temperature (compared with electronic energies) that the electrodes bandwidth
is effectively infinite. Second, we assume that the bosonic environment does
not generate electronic correlations in the electrodes, which is reasonable
for the small electrode coupling that we have here. Within these
approximations, the real-time retarded Green's function,
Eq.~\ref{eq:retarded}, becomes~\cite{jauho94, mahan07}
\begin{equation}
G_{DNA}\left(t\right)=-i\Theta\left(t\right)e^{-i\tilde{E}_{0}t}e^{-\gamma
t}e^{-\phi\left(t\right)}.\end{equation}
 This Green's function includes the coupling to the electrodes through
the factor $e^{-\gamma t}$, where $\gamma$ is the coupling strength
to both electrodes, and includes the coupling to the bosons through
the factor $e^{-\phi\left(t\right)}$ and the renormalized energy
$\tilde{E}_{0}$. The bosonic term is \begin{equation}
\phi\left(t\right)=\sum_{k}\frac{\left|g_{k}\right|^{2}}{\omega_{k}^{2}}\left[n_{k}\left(1-e^{i\omega_{k}t}\right)+\left(n_{k}+1\right)\left(1-e^{-i\omega_{k}t}\right)\right],\end{equation}
where $n_{k}=1/\left(\exp\left(\beta\omega_{k}\right)-1\right)$ is
the equilibrium occupation of mode $k$ at inverse temperature
$\beta$. So long as the temperature is large compared to the boson
cutoff frequency, $\omega_{c}$, then $n_{k}\approx1/\beta\omega_{k}$
and $n_{k}\approx n_{k}+1$, thus \begin{equation}
\phi\left(t\right)\approx\sum_{k}\frac{2\left|g_{k}\right|^{2}}{\beta\omega_{k}^{3}}\left(1-\cos\omega_{k}t\right).\end{equation}
In terms of the spectral function,
$J\left(\omega\right)=\sum_{k}\left|g_{k}\right|^{2}\delta\left(\omega-\omega_{k}\right)$,
\begin{equation}
\phi\approx\int_{0}^{\omega_{c}}\frac{2J\left(\omega\right)}{\beta\omega^{3}}\left(1-\cos\omega
t\right)d\omega.\end{equation} Similarly, the renormalized energy
state is \begin{equation}
\tilde{E}_{0}=E_{0}+\int_{0}^{\omega_{c}}\frac{J\left(\omega\right)}{\omega}d\omega.\end{equation}

For an ohmic boson bath~\cite{garg85},
$J\left(\omega\right)=\alpha\omega$ for $\omega<\omega_{c}$. At high
temperature with respect to its cutoff frequency,
$\phi\left(t\right)\approx\eta t$, where $\eta=\alpha\pi/\beta$, and
$\tilde{E}_{0}=E_{0}+\alpha\omega_{c}$. Generally $\omega_{c}$ is
quite small compared to molecular energies, we thus ignore the
energy shift, which is valid except when the noise strength is very
large. This gives \begin{equation}
G_{DNA}(E)=\frac{1}{E-E_{0}+i\gamma+i\eta},\label{eq:Gapprox}\end{equation}
 for the retarded Green's function. In this work, $\eta=\hbar/\tau_{n}=0,6.6\times10^{-3},6.6\times10^{-2},6.6\times10^{-1},6.6$
eV for the timescales considered. For an interacting junction as
given by the Hamiltonian in Eq. \ref{eq:Ham}, the current is given
by (using Eq. 4.114 in Ref. \cite{diventra08}) \begin{equation}
I\left(\eta\right)=\frac{2e^{2}V}{h}\left[\frac{\gamma^{2}}{E_{0}^{2}+\left(\gamma+\eta\right)^{2}}+\frac{\eta\gamma}{E_{0}^{2}+\left(\gamma+\eta\right)^{2}}\right].\label{eq:ModCurr}\end{equation}

The first and second terms represent precisely the first and second
contribution in Eq.~\ref{eq:transmission}, respectively. However, as
we have previously discussed, within this model calculation, the
liquid environment is allowed to form coherent interactions with the
current-carrying electrons inside the junction. This results in the second term giving
rise to orders of magnitude increase in the total current to values
that are unlikely in the present setting. In the junction, one has
to consider also that the bosonic environment scatters the
current-carrying electrons in all directions, including along the
pore channel where they can be collected into the liquid. This
effect is exacerbated by the fact that the environment both
carries some longitudinal momentum and can act as a sink for
electrons as well, due to the longitudinal bias. Therefore, on
physical grounds, we assume that such processes occur which provide
only the first contribution to the current in Eq.
\ref{eq:transmission}. Again, this is equivalent to assuming a two-probe noise model, as we have discussed previously. 
Under this assumption and for $\gamma\ll
E_{0}$, the expression in Eq. \ref{eq:ModCurr} becomes
\begin{equation}
I(\eta)\approx\frac{2e^{2}V}{h}\frac{\gamma^{2}}{E_{0}^{2}+\eta^{2}},\label{currapprox}\end{equation}
 i.e., the current for just a single structural distortion for linear
response and weak coupling, and in the absence of inelastic processes
that enhance the current. Note, that irrespective of this
approximation, our main conclusions would be qualitatively
unchanged.

We know from above that the current acquires a distribution when structural
distortions of the DNA are taken into account. Under the assumptions
that went into Eq.~\ref{eq:Gapprox} we can introduce these structural
distortions by allowing $E_{0}$ or $\gamma$ to acquire distributions.
From Fig.~\ref{fig:DistributionsA}, it is clear that the current
distributions on a logarithmic scale can be approximated as a Gaussian
when no noise is present, which indicates that the coupling to the
electrodes is controlling these distributions, as only the coupling
fluctuates on an exponential scale. By assuming the coupling to both
electrodes is identical, we miss structural distortions which bring
the base into closer proximity to one electrode and farther from the
other. However, this is unlikely to affect the essential physics.

\begin{figure}
\centering
\includegraphics[width=8.66cm]{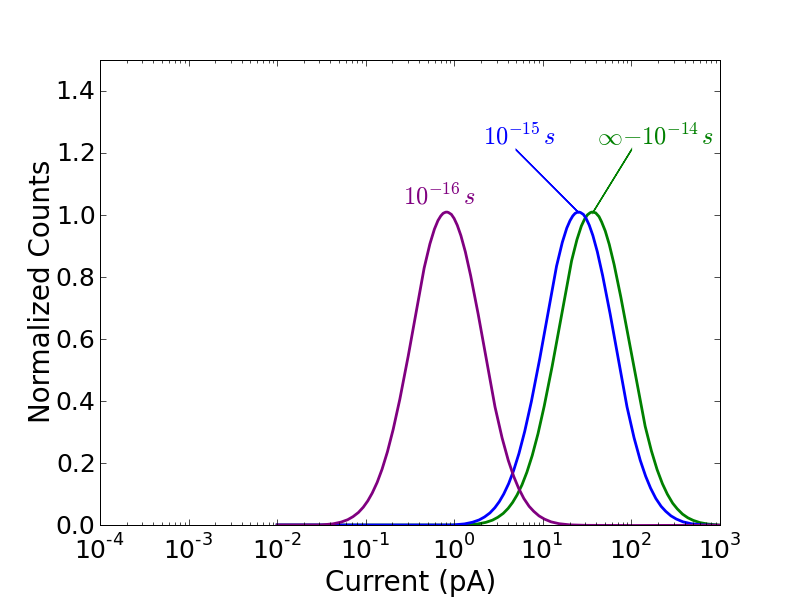}
\caption{Current distributions of a model system for the Adenine
nucleotide represented by a single energy level $E_{0}$. The current
distribution on a logarithmic scale is taken to be Gaussian in
similarity to Fig.~\ref{fig:DistributionsA} for no noise. As noise
is turned on, at first the distribution does not change at all, but
around $\eta\approx E_{0}$, where $\eta=\hbar/\tau_{n}$ measures the
strength of the noise, the distribution starts to shift. At larger
$\eta$, the peak of distribution shifts to lower values as
$\eta^{-2}$. The off-resonant tunneling, indicated by large $E_{0}$
as measured from the Fermi level, ``protects'' the current
distributions from noise.}
\label{fig:analytic}
\end{figure}

\begin{figure}
\centering
\includegraphics[width=8.66cm]{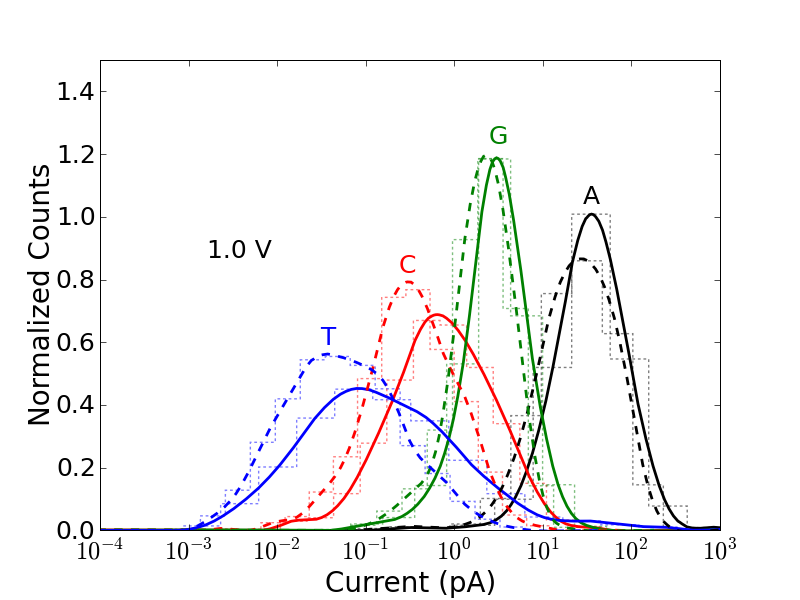} \includegraphics[width=8.66cm]{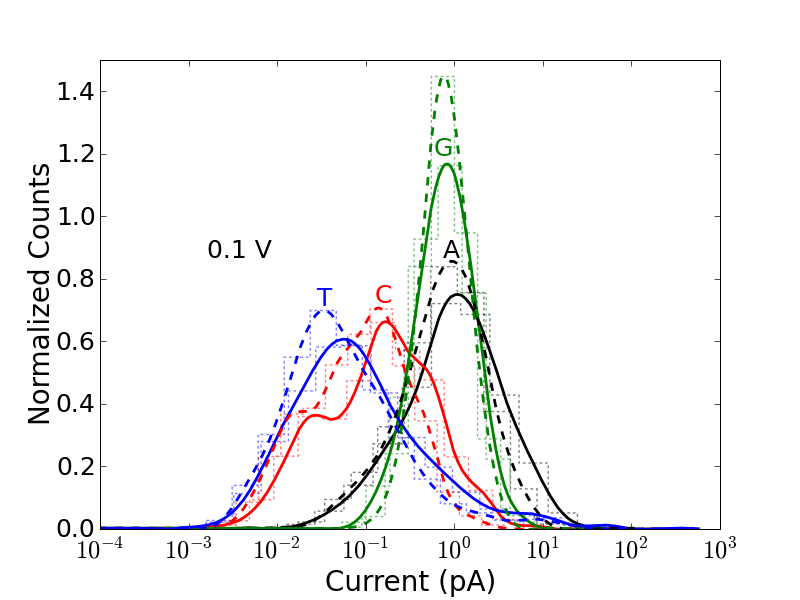}
\caption{Normalized current distributions for the four nucleotides
at a transverse bias voltage of 1.0 V (top) and 0.1 V (bottom). The
solid lines correspond to an infinite noise timescale (no noise) and
the dark dashed lines represent the distributions for
$\tau_{n}=10^{-15}\,$s, with the light dashed lines representing the
bins used to produce the distributions.}
\label{fig:Distributions1V}
\end{figure}

Now, let us calculate the distribution of $\gamma$'s using the curve
in Fig.~\ref{fig:DistributionsA} with no noise. Using the fact that
the current distribution on a logarithmic scale is approximately a
Gaussian, and that we are in a weak coupling regime ($\gamma\ll E_{0}$),
$\ln\gamma/\gamma_{m}$, where $\gamma_{m}$ is the maximum likelihood
coupling strength, should also follow a Gaussian distribution, \begin{equation}
p(\ln\gamma/\gamma_{m})=\frac{1}{\sqrt{2\pi\sigma_{\gamma}^{2}}}\exp\left\{ -\frac{(\ln\gamma/\gamma_{m})^{2}}{2\sigma_{\gamma}^{2}}\right\} ,\end{equation}
 with the standard deviation $\sigma_{\gamma}=\sigma_{I}/2\approx0.45$,
where $\sigma_{I}$ is the standard deviation of $\ln I/I_{m}$ with
$\eta=0$ and $I_{m}$ the maximum value. (Note that the relatively
small standard deviation of the current distributions $\ln I/I_{m}$,
as seen in Figs.~\ref{fig:DistributionsA} and~\ref{fig:Distributions1V},
is a result of the control exerted by the transverse field~\cite{lagerqvist06,lagerqvist07}.
In the absence of such control the current distributions span several
orders of magnitude and have considerable overlap~\cite{lagerqvist07b}.)
The maximum, $\gamma_{m}$, appears at $6.8\times10^{-4}$ eV, when
$E_{0}=1\,$eV, which is approximately the energy separation of Adenine's
LUMO from gold's Fermi level~\cite{zwolak08}. We assume that the
standard deviation of $\ln\gamma/\gamma_{m}$ does not change when
we turn on the noise. The resulting current distributions are plotted
in Fig.~\ref{fig:analytic}.

Although we assume in our model that the distributions stay Gaussian
with the same standard deviation no matter what the noise strength,
our model explains the key features found in our numerical
simulations. The fact that the molecular energy levels are far away
from the electrode Fermi level ``protects'' the distributions from
this type of noise. This is represented by the term
$(E_{0}^{2}+\eta^{2})^{-1}$ in the current (Eq.~\ref{currapprox}).
The other features that appear, such as increased broadening and
eventual multiple peak development, are not explained by our simple
model. These are due to multiple energy levels, $E_{i}$, of the
fluctuating nucleotide junction, contributing to transport. The
contribution from each reaches its turning point, $\eta\approx
E_{i}$, at a different value of $\eta$ and thus the single peak
broadens and develops into multiple peaks.

For the remainder of this paper we examine the role of transverse
bias on the distributions for two different noise strengths (i.e.,
no noise and a timescale of $\tau_{n}=10^{-15}\,$s). The results for
the cases of 0.1 V and 1.0 V transverse biases are presented in
Fig.~\ref{fig:Distributions1V}. Previous work has shown that the
transverse bias has a nonlinear effect on the mean of the
distribution~\cite{lagerqvist07}. This is due to both a pulling
effect of the backbone toward one electrode as the field is
increased with consequent alignment of the base toward the other
electrode, and the steric effect of the alignment of the backbone
with one of the electrodes. Therefore, while one can expect the mean
current to be shifted to lower values with lower bias, the degree to
which this occurs is not easy to determine {\em a priori}. This is
especially true with the smaller base T. For this base, one cannot
always expect perfect alignment at all times with the electrodes
even in the presence of a stabilizing transverse field, further
emphasizing the statistical nature of this problem. These effects
can be seen in Fig.~\ref{fig:Distributions1V}. In addition, one can
see that all of the distributions are shifted slightly to lower
current values due to noise, corresponding to an overall lowering of
the current magnitude. However, the distributions themselves are
very similar to the case of an infinite timescale (zero noise).

\subsection{Conclusion}

In conclusion, we have presented results combining molecular
dynamics simulations with quantum mechanical current calculations
including a model of noise generated by the complex liquid
environment in which the DNA translocation and interrogation takes
place. We have shown that for reasonable timescales, e.g., down to
$10^{-15}\,$s, the noise considered here will likely not affect the distinguishability
of the current distributions obtained from measuring the transverse
electronic current of the different DNA nucleotides. At extremely
fast timescales, below $10^{-15}\,$s, the distributions are
significantly altered, but this is beyond physically reasonable
times for the experimental system we are considering. We have also
proposed a simple model system which provides insight into the
physical mechanisms of noise and why the current distributions are
protected. This is due to the off-resonant nature of tunneling
through the nucleotides and thus it is likely to be a general
property of transport in organic molecules. While the distributions
are only mildly affected, we have shown that the type of noise we
consider can potentially alter a single current count significantly,
further supporting the notion that only a statistical study of the
transverse currents can potentially distinguish the nucleotides. We
finally note that while our study is done for a nanopore geometry,
the results are applicable to other types of sequencing devices as
well, such as the nanochannels of Refs.~\cite{liang08,maleki09} used
in transverse electronic measurements.
\begin{acknowledgments}
We thank Yonatan Dubi and Johan Lagerqvist for useful discussions.
This research is supported by the NIH-National Human Genome Research
Institute and by the U.S. Department of Energy through the LANL/LDRD
Program.
\end{acknowledgments}

\bibliographystyle{biophysj}
\bibliography{dephasing}

\end{document}